\documentclass[%
 reprint,
 superscriptaddress,
showpacs,
 amsmath,amssymb,
 aps,
]{revtex4-1}
\usepackage{graphicx}% Include figure files
\usepackage{dcolumn}% Align table columns on decimal point
\usepackage{bm}% bold math
\usepackage{color}
\usepackage[usenames,dvipsnames]{xcolor}
\begin{document}

\preprint{APS/123-QED}
\title{Phase matching effects in high harmonic generation at the nanometer scale}

\author{M. Blanco}
\affiliation{Grupo de investigaci\'on Photonics4Life, \'Area de \'Optica, Departamento de F\'isica Aplicada, Universidade de Santiago de Compostela, Campus Vida s/n, Santiago de Compostela, Spain}
\author{C. Hern\'andez-Garc\'ia}
\affiliation{Grupo de Investigaci\'on en Aplicaciones del L\'aser y Fot\'onica, Departamento de F\'isica Aplicada, University of Salamanca, E-37008, Salamanca, Spain}
\author{A. Chac\'on}
\affiliation{ICFO - Institut de Ci\`encies Fot\`oniques, The Barcelona Institute of Science and Technology, 08860 Castelldefels (Barcelona), Spain}
\author{M. Lewenstein}
\affiliation{ICFO - Institut de Ci\`encies Fot\`oniques, The Barcelona Institute of Science and Technology, 08860 Castelldefels (Barcelona), Spain}
\affiliation{ICREA, Lluis Companys 23, 08010 Barcelona, Spain}
\author{M.T. Flores-Arias}
\affiliation{Grupo de investigaci\'on Photonics4Life, \'Area de \'Optica, Departamento de F\'isica Aplicada, Universidade de Santiago de Compostela, Campus Vida s/n, Santiago de Compostela, Spain}
\author{L. Plaja}
\affiliation{Grupo de Investigaci\'on en Aplicaciones del L\'aser y Fot\'onica, Departamento de F\'isica Aplicada, University of Salamanca, E-37008, Salamanca, Spain}

\date{\today}% It is always \today, today,
             %  but any date may be explicitly specified

\begin{abstract}
Plasmon resonances are known to amplify the electromagnetic fields near metallic nanostructures, providing a promising scheme to generate extreme-ultraviolet harmonics using low power drivings. During high-order harmonic generation (HHG), the driving and harmonic fields accumulate a phase difference as they propagate through the target.  In a typical set-up --a laser focused into a gas jet-- the propagation distances amount to several wavelengths, and the cumulative phase-mismatch affects strongly the efficiency and properties of the harmonic emission. In contrast, HHG in metallic nanostructures is considered to overcome these limitations, as the common sources of phase mismatch --optical density and focusing geometry--  are negligible for subwavelength propagation distances. We demonstrate that phase matching still plays a relevant role in HHG from nanostructures due to the non-perturbative character of HHG, that links the harmonic phase to the intensity distribution of the driving field. Our computations show that widely used applications of phase matching control, such as quantum path selection and the increase of contrast in attosecond pulse generation, are also feasible at the nanoscale. 
\end{abstract}

\maketitle

\section{Introduction}

%\section{Introduction}
HHG driven by strong laser fields is demonstrated as a reliable scheme for the generation of short-wavelength coherent radiation. Its extraordinary characteristics reserve HHG
%harmonic generation 
its own niche among other sources of high-frequency radiation.  Recently, HHG has been boosted into the soft x-ray spectral region using mid-IR drivings and high-pressure gases in hollow  wave-guides \cite{popmintchev12}. Most notably, HHG also conveys the production of short-wavelength radiation in the form of ultrashort pulses, with subfemtosecond durations \cite{paul01}. Different techniques allow to generate isolated attosecond pulses, even using multicycle drivers (see Ref. \cite{chini14} and references therein). Therefore HHG is a fundamental tool not only for spectroscopy, but also for time-dependent measurements at the attosecond scale \cite{plaja13, Krausz2009}. 

The requirement of a strong driving links HHG to the limitations of intense-laser amplification schemes. The widely used Chirped-Pulse Amplification, reduces by two or three orders of magnitude the oscillator's repetition rate, limiting the energy of the time-integrated harmonic signal. As an alternative, it has been shown that the oscillator's electric field can be amplified enough to drive HHG, when focused into a target enclosed into a metallic nanostructure \cite{kim08,sivis12,pfullmann13,Han16,vampa17}, and the HHG signal has been reported for solid targets \cite{Han16,vampa17}. The laser field can induce resonant plasmon oscillations in metallic nano-gaps, with local field intensities amplified by more than two orders of magnitude. 
A first theoretical investigation of HHG in bow-tie antennas and nanocones reported the extension of the plateau's cutoff frequency \cite{husakou11}. The generation of isolated attosecond pulses was subsequently studied theoretically in coupled-ellipsoids nanostructures \cite{stebbings11}, and the application of polarization gating for the same end in bow-tie structures has been addressed in \cite{stebbings11}. HHG and attosecond pulse generation has also been studied in nano-structured metallic funnel-waveguides \cite{choi12}. Plasmonic-field amplification also conveys phenomenology in multiphoton ionization and delayed photoemision  \cite{ciappina16}. In the experiments  \cite{kim08,sivis12,pfullmann13}, bow-tie shaped elements were etched in gold films evaporated on sapphire substrates and submitted to  $\leq$10 fs,  800 nm oscillator pulses with focused peak intensity of $\sim$ 10$^{11}$ W/cm$^2$, at 78 MHz repetition rate. The bow-tie gaps are filled with Ar or Xe, exiting a gas nozzle at 115 to 375 Torr pressures. Under these conditions, and for bow-tie designs similar to those in the present study, the plasmonic resonance enhances the driving field in 20 to 40 db \cite{kim08}. It should be noted, however, that the field enhancement is located at the antenna's gap, and not at the substrate, where the intensity remains below the damage threshold. Gold nanoantennas can tolerate high field amplitudes before being damaged, depending on the particular conditions of resonance, reaching intensities up to 10$^{14}$ W/cm$^2$ for antenna's lengths of 140-175 nm and $<$ 10 fs pulse lengths \cite{pfullmann13}. This corresponds to an energy deposition of the order of that reported for gold nanowires attached to silica \cite{summers14}. Despite these promising characteristics, the small interaction volume in comparison with that of the conventional HHG experiments at higher intensities, reduces the efficiency of the target volume emission in 6 to 8 orders of magnitude \cite{sivis12,kim12}. Also, resonant-enhanced atomic line emission \cite{sivis12,sivis13} can mask the harmonic signal, making it difficult to distinguish between HHG and atomic fluorescence. Very recently, high-harmonics have been efficiently produced from nanostructures on sapphire \cite{Han16} and silicon \cite{vampa17}.

%In this situation, resonant-enhanced atomic line emission \cite{sivis12,sivis13} can mask the harmonic signal, making it difficult to distiguish among the HHG (coherent) emission and the atomic fluorescent (incoherent) emissions. Further work points out that other geometries, as hollow metal waveguides \cite{choi12}, rough metallic surfaces \cite{kim11} may be more adequate. Also the increase of the number of radiating dipoles, using higher pressures \cite{sivis13}.

The harmonic emission yield from an extended target results from the coherent superposition of contributions from different points in space. The coherence length of the harmonic propagation is affected  by the focusing geometry, neutral atom and free charge densities \cite{gaarde08}, as well as group velocity matching \cite{hernandez16}. An appropriate spatial matching of the harmonic phase can be used as an additional knob to control the properties of the integrated target's emission. Applications of phase matching are not limited to the optimization of the harmonic yield, it is also used as a method to generate isolated attosecond pulses \cite{hernandez16}, or to eliminate the effect of the carrier-envelope shifts when using few-cycle pulses \cite{hernandez15}, among others \cite{gaarde08}. 

The most common sources of phase shifts --focusing, neutrals and plasma-- have a marginal role in HHG in gases when the propagation distances are shorter than the driver's wavelength (see methods in \cite{sivis13}). Considering this, phase matching effects should be residual in HHG from nanostructures, even with the relative high pressures used in the experiments mentioned above. 

Counterintuitively, we demonstrate in this paper that phase matching plays a relevant role even at the nanometric scale, due to the non-perturbative character of HHG. We present full calculations of HHG in bow-tie nano antennas filled with argon, which includes harmonic generation, propagation and the ab-initio computation of the resonant enhancement of the incident field. The plasmon-enhanced  electric field enclosed in these nanostructures is strongly inhomogeneous. We show that the intensity gradients are translated to spatial variations of the harmonic intrinsic phase, affecting phase matching even in nanometric propagations. 
%{\color{red}Such intensity gradients are should be found in the propagation direction in the experimental set-ups  \cite{kim08,sivis13}, where we observe trajectory selection by phase matching. On the other hand, we also show that the transversal intensity distribution is homogeneous enough to allow the coherent build up of the harmonic signal in antennas, with gap distances larger than those used in the mentioned experiments.}
    
The mechanism underlying HHG is non-perturbative, requiring an intense driving field to tunnel-ionize electrons from the parent atoms. Electrons are released into the continuum in the form of  attosecond wavepackets, that are accelerated by the driving field, to subsequently recollide with the parent ion \cite{schafer93,corkum93,lewenstein94}. The harmonic emission results from the conversion of the electron's kinetic energy into high-frequency radiation, during these ultrafast recollisions. Typically, every half-cycle of the driving field, any particular harmonic is generated  by two different electron trajectories, that rescatter with the same kinetic energy. These trajectories are usually referred as {\em long} and {\em short}, according to the time the electron spends in the continuum before recollision. As a result of this non-perturbative dynamics, the electron's wavepacket acquires a phase proportional to its action. This quantum phase is translated to the harmonic radiation as an {\em intrinsic  phase}, proportional to the product of the field intensity and the electron's excursion time in the continuum \cite{lewestein95}. Thus, the intrinsic phase becomes an additional factor of phase matching in HHG, and spatial phase shifts should be expected as the focused driving has an inhomogeneous intensity \cite{hernandez13}. The intrinsic phase has been a key element for the observation, control and manipulation of the efficiency of {\em short} and {\em long} trajectories, and to sculpt the properties of the harmonic emission \cite{salieres01,mairesse03,zair08,hernandez12,hernandez13Z,hernandez15Q}.

\section{Interaction scheme and theoretical methods}

\subsection{Interaction scheme: HHG in a bow-tie nano antenna}

\begin{figure}[ht]
\centering\includegraphics[width=\linewidth]{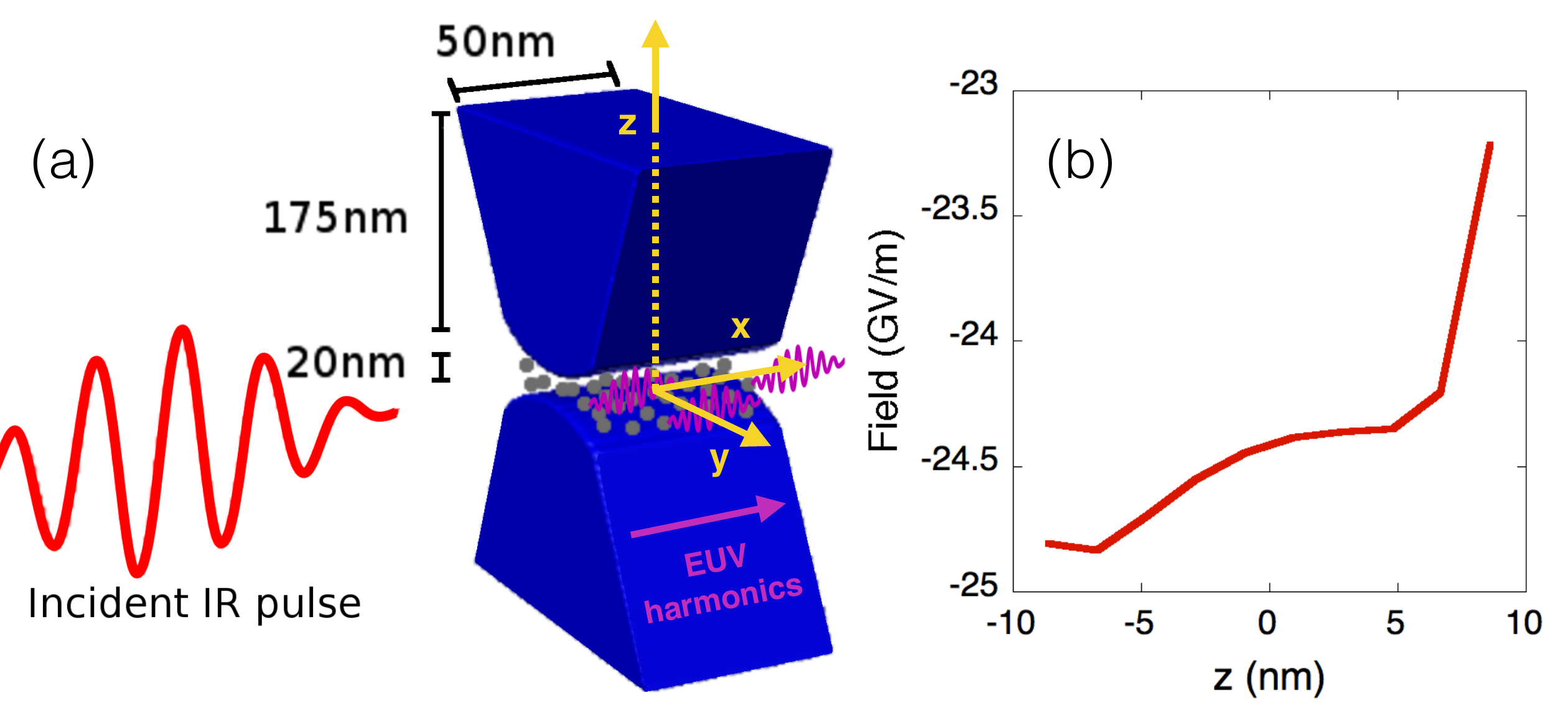}
%\vskip -1 cm 
\caption{(a) Interaction scheme considered in this paper: a few-cycle NIR beam is focused into a gold bow-tie nano antenna. Harmonics are generated from argon gas by the enhanced driving field at the antenna's gap. (b) Field amplitude profile across the gap, at the antenna's center, induced at the peak of the driving field at the time $t_{max}$.}
\label{fig:set-up} 
\end{figure}

Fig. \ref{fig:set-up}a shows a scheme of the system considered in this paper: an 800 nm laser pulse, with sinus-squared 4.7 fs FWHM envelope, is aimed to an array of bow-tie shaped nanoantennas (one of them shown in the figure), whose gaps are filled with Ar gas. The driving field's peak intensity, 6.7 $\times$ 10$^{11}$ W/cm$^2$ (field amplitude 2.25 GV/m), obtained at the simulation time $t_{max} = 8~$fs, %{\color{red} field $4.38 \times 10^{-3} au = 2.25 \times 10^9$ V/m}, 
is chosen as a safe value below the damage threshold \cite{pfullmann13}. The field is linearly polarized along the $z$ axis. The antenna dimensions are 50 nm (thickness), 175 nm (half length) and 20 nm (gap), similar to those used in previous experiments \cite{kim08,sivis13,pfullmann13}. %{\color{red} kim l=175, d=20, t=50, thickness=, pfullmann l=140, d=20, t=50, Sivis, l=200-240, d=20, t=50-90}. 
To compute the field at the antenna's gap we use the particle-in-cell (PIC) code OSIRIS \cite{fonseca02,fonseca2008,fonseca2013}. The integration volume is a $150 \times 300 \times 470$ nm box that encloses completely the antenna design depicted in Fig. \ref{fig:set-up}. We have used a spatial resolution of $1.9$ nm and a time step of $3.2$ as. 

%\begin{verbatim}

Fig. \ref{fig:set-up}b shows the plasmon-enhanced field amplitude profile inside the antenna's gap, along the z axis, at the moment when it reaches its peak intensity ($t_{max}$). From this graphic, we estimate a plasmon-enhancement factor 11 for the field amplitude, > 20 dB enhancement in intensity, in agreement  with \cite{kim08}. Note that the field intensity distribution inside the gap is inhomogeneous. Previous studies have shown that for strong spatial inhomogeneities, the harmonic spectrum radiated by a single atom is modified \cite{ciappina12}, extending to frequencies beyond the spectral cut-off. Our integration of the time-dependent Schr\"odinger equation indicate that this effect is negligible for the interaction parameters considered here (see Appendix). 

\subsection{Theoretical methods: HHG combined with Particle-In-Cell}

HHG and propagation is computed at the antena's gap volume using the electromagnetic field propagator \cite{hernandez10} for a sufficiently large sample of individual atoms  ($10^5$). We compute the harmonic propagation using the electromagnetic field propagator \cite{hernandez10}. The target (gas cell or gas jet) is discretized into a set of elementary radiating volumes centered at ${\bf r}_j$, and propagate the emitted field {${\bf E}_j({\bf r}_j,t)$ to the far-field detector, located at ${\bf r}_d$,

\begin{equation}
{\bf E}_j ({\bf r}_d, t)={q_j  {\bf s}_d\over  c^2| {\bf r}_d-{\bf r}_j |}  \times \left[ {\bf s}_d \times {\bf a}_j \left(t-{| {\bf r}_d-{\bf r}_j | \over c}\right) \right]
\label{eq:E_discrete}
\end{equation}

\noindent where ${\bf a}_j$ is the dipole acceleration, $q_j$ is the charge of the electron, ${\bf s}_d$ is the unitary vector pointing to the detector, and ${\bf r}_d$ and  ${\bf r}_j$ are the position vectors of the detector and of the elementary radiator $j$, respectively. 
Equation (\ref{eq:E_discrete}) assumes that the harmonic radiation propagates with the vacuum phase velocity, which is a reasonable assumption for high-order harmonics. Finally the total field at the detector is computed as the coherent addition of these elementary contributions. Propagation effects in the fundamental field, such as the production of free charges, the refractive index of the neutrals, the group velocity walk-off \cite{hernandez16}, as well as absorption in the propagation of the harmonics, are included, although they are negligible in gas targets for sub-wavelength propagation distances. 

In the case of intense fields, the computation of the dynamics of the elementary radiators is not trivial, as the interaction is non-perturbative. Due to the large number of radiators, using the exact numerical integration of the TDSE becomes extremely expensive, specially for mid-infrared driving fields. Therefore, the use of simplified models is required. For these intense fields, the S-matrix approaches combined with the Strong-Field Approximation (SFA) \cite{keldysh64,faisal73,reiss80} are demonstrated to retain most of the features of the HHG \cite{lewenstein94,becker97}. We have recently developed an extension of the standard SFA (SFA+), where the acceleration of the $j$ radiator (${\bf a}_j$) is found from two contributions, ${\bf a}_{b,j}$ and ${\bf a}_{d,j}$, the first being the standard SFA expression, and the later being a correction due to the instantaneous dressing of the ground state during the electron's recollision. Our method computes the dipole acceleration directly from the superposition of the contributions of each Volkov wave. Each contribution can be integrated separately as an ordinary 1D equation, leading to a very efficient algorithm without resorting to the saddle-point approximation \cite{perez09,perez11}. Our method combining high harmonic generation and propagation has shown excellent agreement with HHG experiments with conventional set-ups, where phase matching plays a relevant role \cite{popmintchev12,hernandez13,hernandez15}. For the pressures ($<$ 375 Torr) and nanometric optical paths considered here, the influence of the neutral and free carriers on the driving field is negligible \cite{sivis13}, consequently the harmonic intensity signal follows a power-two scaling with the number of radiators, typical of a coherent build-up process. To obtain a faithful comparison of the phase matching conditions at different interaction volumes, the results of our calculations have been normalized to the same number of radiators.

We note that, despite quantum SFA models are derived for homogenous fields, results in the Appendix show that the field inhomogeneity is irrelevant at the single-atom level, and therefore SFA can be used our simulations. 

The plasmon resonance is calculated using the Particle-In-Cell (PIC) code OSIRIS \cite{fonseca02,fonseca2008,fonseca2013}. PIC codes are an useful tool to compute the interaction between fields and charged particles and the dynamics of the interaction between charged particles within plasmas \cite{Birdsall:1991}. These codes solve in a regular grid, for each timestep, the full set of Maxwell's equations to compute the evolution of the fields, and Lorentz's equations to account for the particle motion. Each timestep the particles are moved, current and charge densities are deposited on the grid, and the fields are recalculated. These codes are fully relativistic and suitable for complex situations with several degrees of freedom. They have proven useful to compute laser-matter interaction at the nano- and micro- scales, giving high quality results for laser-plasma based particle accelerators or for HHG in solid targets.

%\begin{verbatim}
\begin{figure}[ht]
%\centering\includegraphics[width=7cm]{opexfig1}
\centering\includegraphics[width=\linewidth]{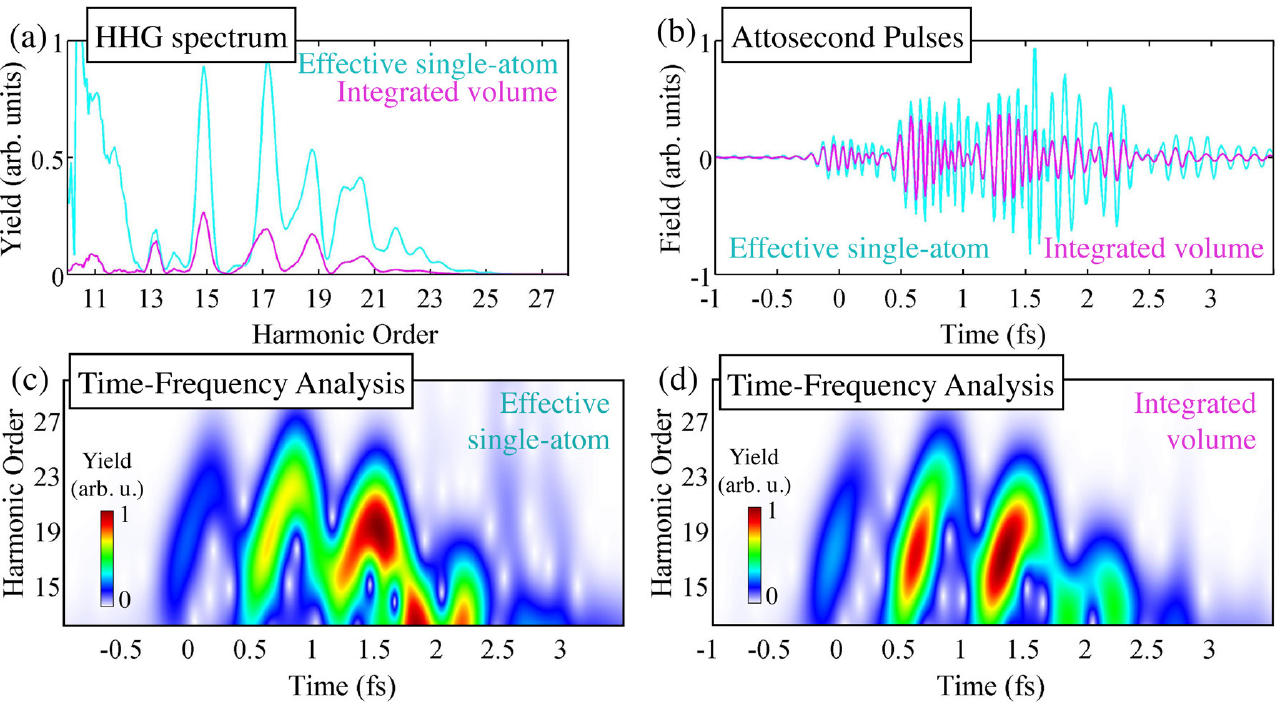} 
\caption{(a) Violet: Harmonic spectrum from the volume-integrated propagated signal
emitted by the argon gas enclosed in the antenna's gap (linear scale), for the antena's design depicted in  Fig. \ref{fig:set-up}a.
. Blue: emission by the same number of atoms positioned at the coordinate origin (referred in the text as 
effective single-atom emission).
 (b) EUV field emitted by the antenna, after filtering the harmonic orders below 10 (c) Time-frequency analysis of the emission of the effective single-atom (d) the same for the volume-integrated propagated signal.}
\label{fig:caso_ref} 
\end{figure}

\section{Results}

In Fig. \ref{fig:caso_ref}a we plot in violet the total (volume-integrated) propagated harmonic emission from the gas enclosed in the gap, for the antenna structure depicted in Fig \ref{fig:set-up}a, as collected by a far-field detector located in the x-axis. %for a gas density of 1$\times$10$^{20}$ atoms/cm$^3$. 
%The harmonic spectra is shown in Fig. \ref{fig:caso_ref}a in violet, 
For comparison, the effective single-atom emission is shown in blue. We define the latter as the coherent emission of the same number of atoms in the volume, all located at the coordinate origin (gap's center), therefore propagation effects are excluded. Note that the drop in harmonic efficiency by a factor 4 reflects the fact that  the average intensity in the interaction volume is lower than the intensity at the gap's center --where the effective single-atom emission is computed--, so it can not be considered as a direct proof of phase mismatch degradation. In contrast, panels \ref{fig:caso_ref}b to d provide for a more precise diagnostic of the role of phase matching.

Figs. \ref{fig:caso_ref}c and d plot the time-frequency analysis of the harmonic emission for the effective single-atom and the integrated volume, respectively. The wedge-shaped intensity structures have a well-known origin \cite{antoine95}, corresponding to the pair of electron trajectories that rescatter with the parent ion with the same kinetic energy every half cycle. The first harmonic emission corresponds to the short trajectory (positive slope) and the second to the long one (negative slope). The comparison of Figs. \ref{fig:caso_ref}c and d shows that the contribution of the long trajectories is strongly suppressed in the integrated-volume emission. The elimination of trajectory contributions is a well known consequence of phase matching in HHG, and has a practical interest for attophysics, as the associated attosecond bursts become narrower. This is shown in Fig. \ref{fig:caso_ref}b, where we plot the temporal structure of the higher part of the harmonic spectra shown in \ref{fig:caso_ref}a (harmonic orders $>$ 10). It becomes apparent how the contrast and width of the attosecond pulses is narrower in the integrated volume case. Also, as both trajectory contributions exhibit opposite chirp, the suppression of one type leads to attosecond pulses with simpler chirp, easier to be compensated to near the Fourier limit \cite{Sansone2006}.

%\begin{verbatim}
\begin{figure}[ht]
\centering\includegraphics[width=\linewidth]{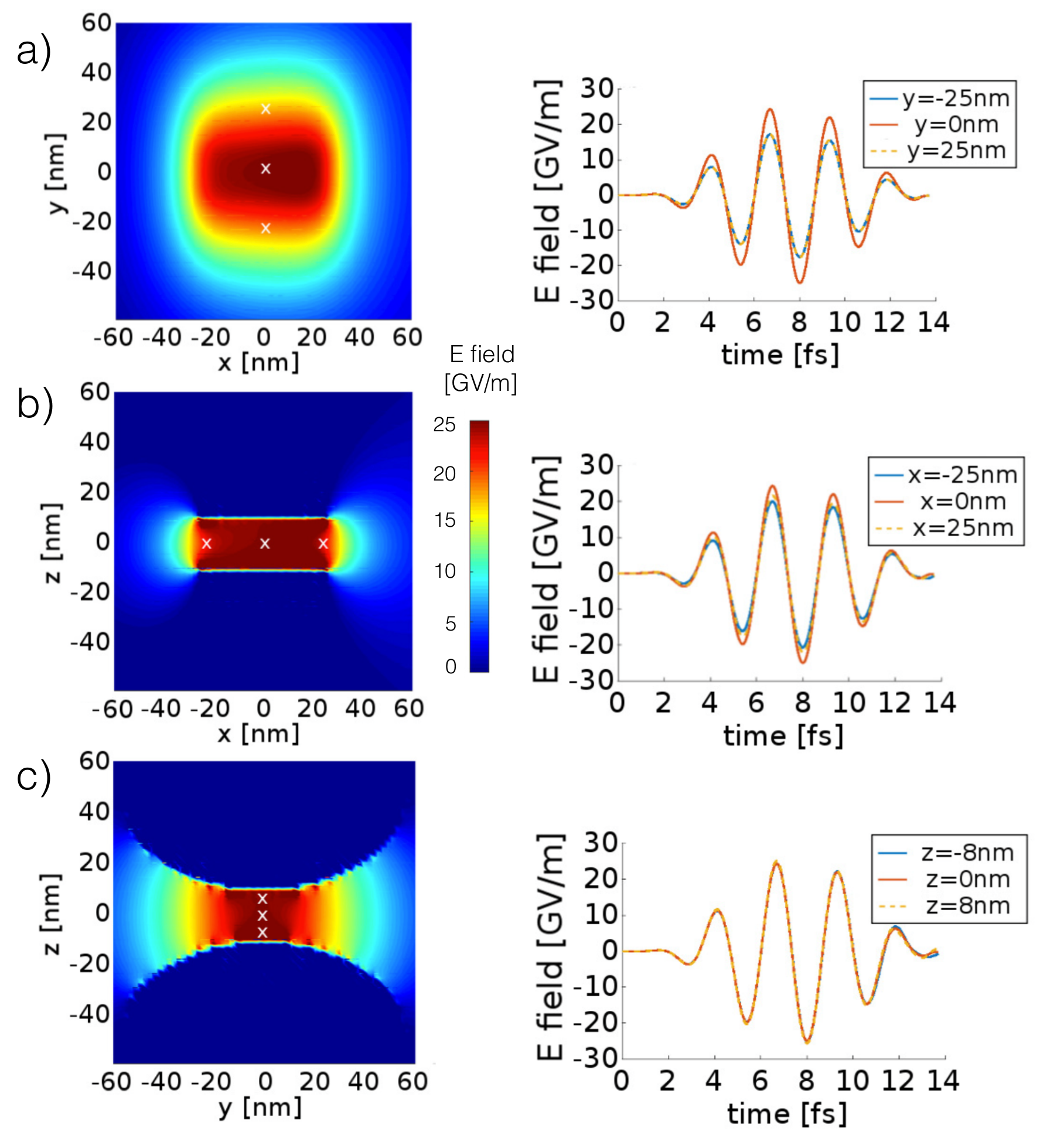} 
\caption{Left column: Absolute value of the field amplitude at the planes (a) $z=0$, (b) $y=0$ and (c) $x=0$ at the time $t_{max}$. Right column: Time-dependent amplitudes at three points in each of the planes (white crosses in the left-column plots).}
\label{fig:distributions}
\end{figure}

% \section{Phase mismatch in nanoscale propagation.}
To gain insight into the nature of this suppression, we plot in the left column of Fig. \ref{fig:distributions} the local field intensity distributions inside the antenna's gap, for the main planes. The intensity distributions are plotted at the time instant when the driving field has its maximum ($t_{max}$). As expected, the antenna structure imposes intensity variations at the nanometer scale, the intensity gradients being more pronounced in the directions transverse to the antenna's gap. On the right we plot the time dependent field amplitudes at three points along the y, x and z directions, respectively. The inspection of these plots shows no appreciable phase shifts of the driving field within the gap's volume, therefore, showing that the nanostructure geometry does not have an effect in the field's phase, as it does in the intensity.

%The inset of (b) shows a small phase shift of the carrier (CEP) when comparing the fields at the entrance and the exit of the gap ($x$ direction). According to the SFA models, this phase shift is mapped to the harmonic field as as multiple of the harmonic order. On the other hand, the insets of (a) and (c) show no relevant phase shift of the field induced by the nanostructure along the $y$ and $z$ axis.

Therefore, our analysis rules out three of the main factors contributing to the phase mismatch of harmonic propagation: on one side, neutrals and plasma contributions are negligible in this subwavelength scale of the propagation, on the other side, the nano-antenna geometry does not have a relevant effect in phase shifting the field. This leaves the intrinsic phase as the main source of phase mismatch. Since  this phase term is proportional to the local intensity, the spatial inhomogeneity shown in  Fig. \ref{fig:distributions} will be translated to phase shifts of the harmonics emitted at different regions of the interaction volume. Note that this remarkably small scale of phase mismatch is induced by the non-perturbative nature of the HHG and, therefore, will not be found in ordinary perturbative harmonic generation.

%\begin{verbatim}
\begin{figure*}[ht]
%\centering\includegraphics[width=7cm]{opexfig1}
\centering\includegraphics[width=0.7\linewidth]{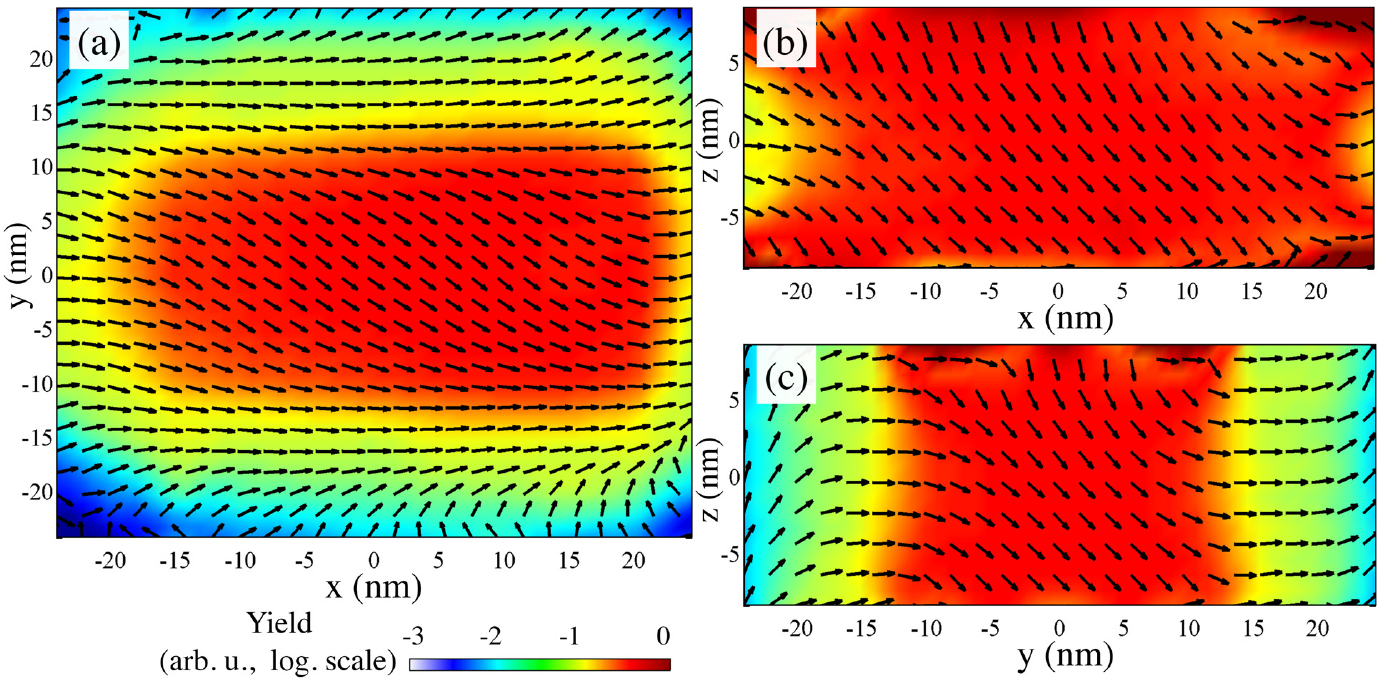} 
\caption{Intensity (color) and phase (arrows) maps of the spatial contributions to the 17th harmonic far field plotted along the planes shown in Fig. \ref{fig:distributions} a to c, respectively. %The color scale represents the intensity of the radiated harmonic and the harmonic phase is represented by the orientation angle of the arrows.
}
\label{fig:arrows}
\end{figure*}

To demonstrate the spatial shift of the harmonic phase, we show in Figs. \ref{fig:arrows}a to c the spatial maps of the intensity (color background) and phase (arrows) for the 17th harmonic of the spectrum shown in Fig. \ref{fig:caso_ref}a, along the main planes of the interaction volume. The arrow distributions define a coherence length in the medium as the distance between the nearest points with opposite arrow directions, i.e. the distance between two atoms whose harmonic emission interferes destructively. Note that HHG is susceptible to phase shifts in the propagation direction as well as in the transverse dimensions \cite{hernandez13}. In the case depicted in Fig. \ref{fig:arrows}, the interaction volume is too small to include a complete half rotation of the arrows, therefore the coherence lengths are larger than the target's size in any dimension. However, despite there is not a complete destructive interference, it is also evident that there is a spatial phase shift in the harmonic emitted and, therefore, the total yield should be affected by the partial cancellation. It can also be observed that these phase shifts are more pronounced in regions with higher intensity gradients of the driving field (shown in Fig. \ref{fig:distributions}). This demonstrates the connection between the harmonic phase shift and intensity variations of the driving field, a particular feature of the intrinsic phase in HHG. As mentioned above, the intrinsic phase is proportional to the electron's excursion time. Long trajectories have a stronger dependence with the intensity than shorter and, therefore, are more susceptible to  intrinsic-phase mismatch. This explains the selective suppression of the long trajectory contributions, revealed by the time-frequency analysis shown in Fig. \ref{fig:caso_ref}. 

%\begin{verbatim}
\begin{figure*}[ht]
%\centering\includegraphics[width=7cm]{opexfig1}
\centering\includegraphics[width=0.7\linewidth]{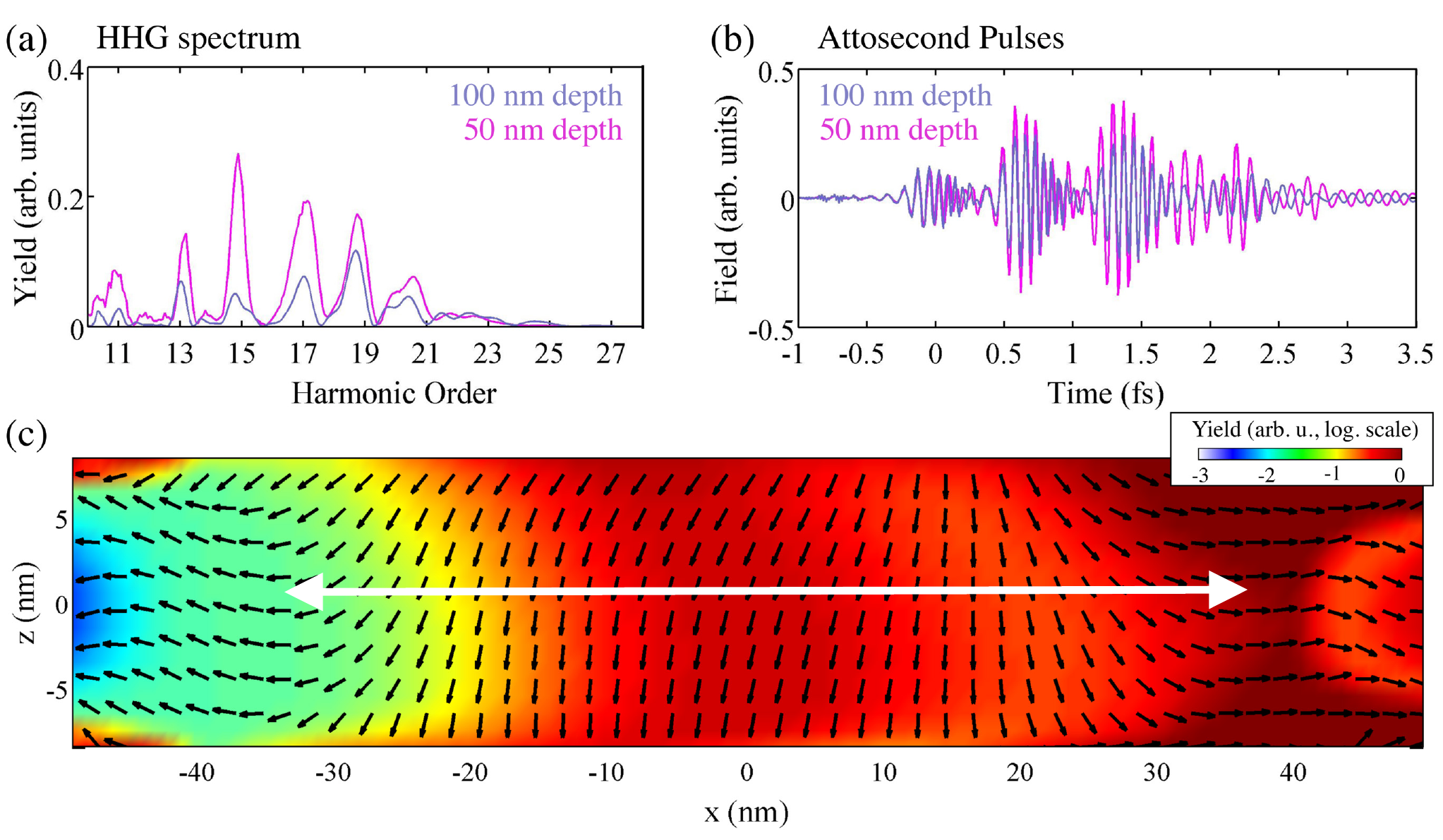} 
\caption{Results for the spectrum emitted by a bow-tie antenna structure with the same geometry as the considered in the main text (Fig. 1, main text) but with its thickness increased by a factor two (100 nm). The driving field parameters are also those used in the main text, except for the driver's intensity, that has been increased by a factor 1.5, to match the peak field amplitude at the center of the gap in both cases. (a) Spectrum of the gap's volume harmonic emission in comparison with the 50 nm thickness case,  shown in Fig. 2 (main text). (b)  Attosecond pulses after selecting the higher frequencies of the harmonic spectra (harmonic orders > 10), in comparison with the 50 nm thickness case. (c) Map of the 17th harmonic intensity (color background) and phases (arrows) for the 100 nm thick antenna, that can be compared with the 50 nm thick case shown in Fig. 4 (main text). The white double arrow indicates the coherence length.}
\label{fig:thickness}
\end{figure*}

\subsection{Harmonic generation in modified antennas}

The modification of the antenna structure affects the plasmon resonance as well as the geometry of the enhanced field. We have explored the effect in HHG of increasing the antenna's thickness and gap dimensions of the original antenna structure, shown in Fig. \ref{fig:set-up}. 

Figure \ref{fig:thickness} depicts the results obtained for a 100 nm thick bow-tie antenna. Our computations show that for this case, the field enhancement in the gap is less than in the 50 nm thick case in the main text. In order to compare properly both situations, we have increased the driving field's intensity in the 100 nm case by a factor 1.5, so the peak field of the amplified field inside the gap remains the same in both calculations. Under this conditions figures \ref{fig:thickness}a and b show that a thicker nano-antenna is less efficient for the harmonic generation and it does not present a clear advantage in the temporal characteristics of the attosecond pulses. The map plotted in figure \ref{fig:thickness}c shows that the interaction region is larger than the coherence length.

%\begin{verbatim}
\begin{figure*}[ht]
%\centering\includegraphics[width=7cm]{opexfig1}
\centering\includegraphics[width=0.7\linewidth]{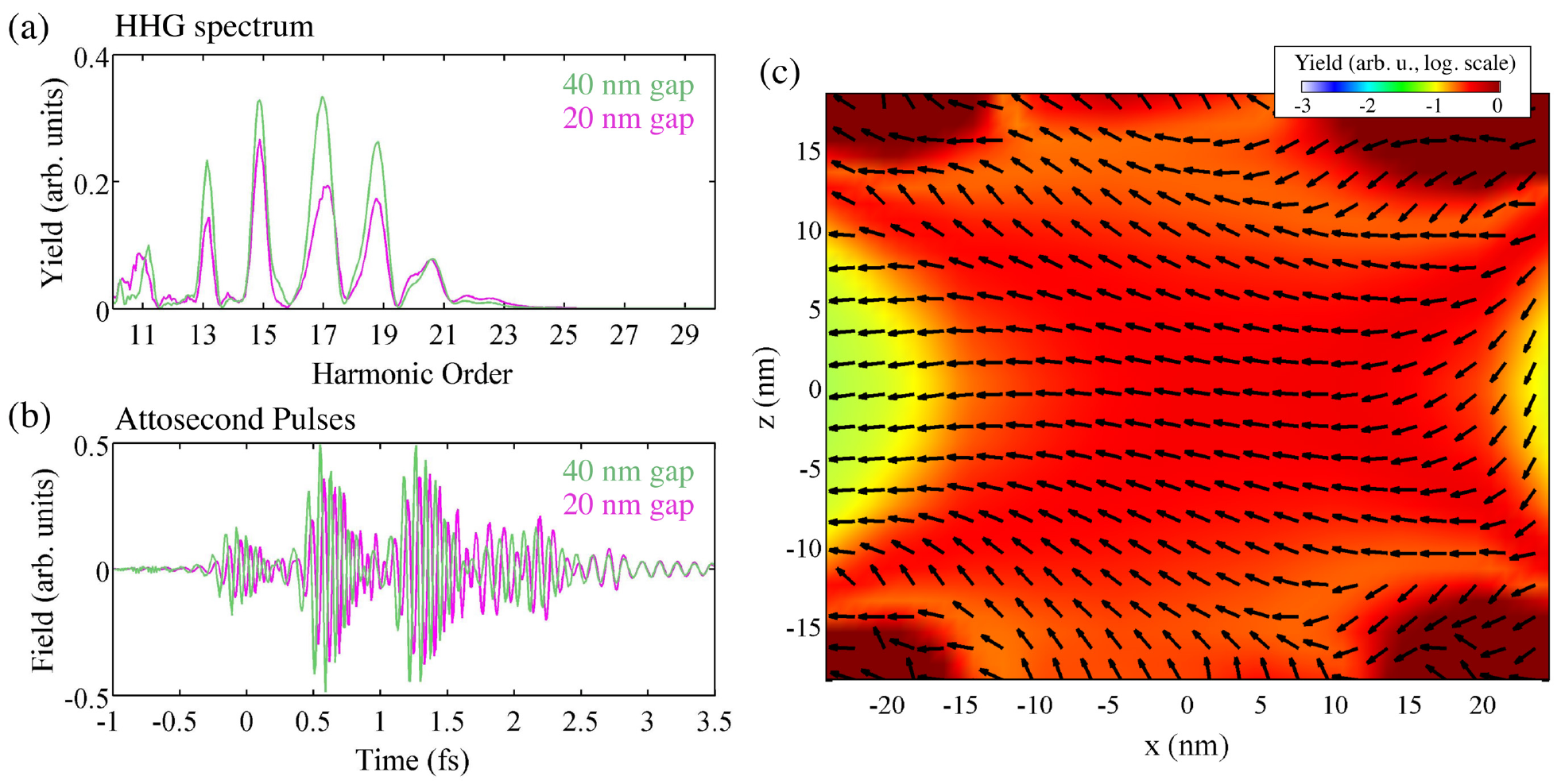} 
\caption{Results for the HHG emitted by a bow-tie antenna structure with the same geometry as the considered in the main text (Fig. 1) but with the gap size increased by a factor two (40 nm). The driving field parameters are also those used in the main text, except for the driver's intensity, that has been increased by a factor 1.77, to match the peak field amplitude in the gap in both cases. (a) Spectrum of the gap's volume harmonic emission in comparison with the 20 nm gap case,  shown in Fig. 2 (main text).(b)  Attosecond pulses after selecting the higher frequencies of the harmonic spectra (harmonic orders > 10), in comparison with the 20 nm gap case. (c) Map of the 17th harmonic intensity (color background) and phases (arrows) for the antenna with 40 nm gap, that can be compared with the 20 nm gap case shown in Fig. \ref{fig:arrows}.} 
\label{fig:gap}
\end{figure*}

In contrast, increasing of the gap distance from 20 nm to 40 nm conveys an enhancement of the efficiency, as shown in Fig. \ref{fig:gap}a , demonstrating that structures with wider gaps are optimal for harmonic generation. In order to compare properly both situations, we have increased the driving field intensity in the 40 nm case by a factor 1.77, such that the peak of the amplified field inside the gap remains the same in both situations. Interestingly, the generated attosecond pulses have a higher contrast, which suggest that the suppression of the long trajectory contributions is more effective for wider gaps. The arrow map  (Fig. \ref{fig:gap}c) shows that since the phase shift along the gap is small, the increase of the gap distance does not include regions with oposite phases in the interaction volume. Therefore wider gaps seem to be more favorable for the efficient harmonic generation, provided that the weaker plasmon resonant enhancement is compensated by the increase of the driving field intensity.

\begin{figure}[ht]
\centering\includegraphics[width=\linewidth]{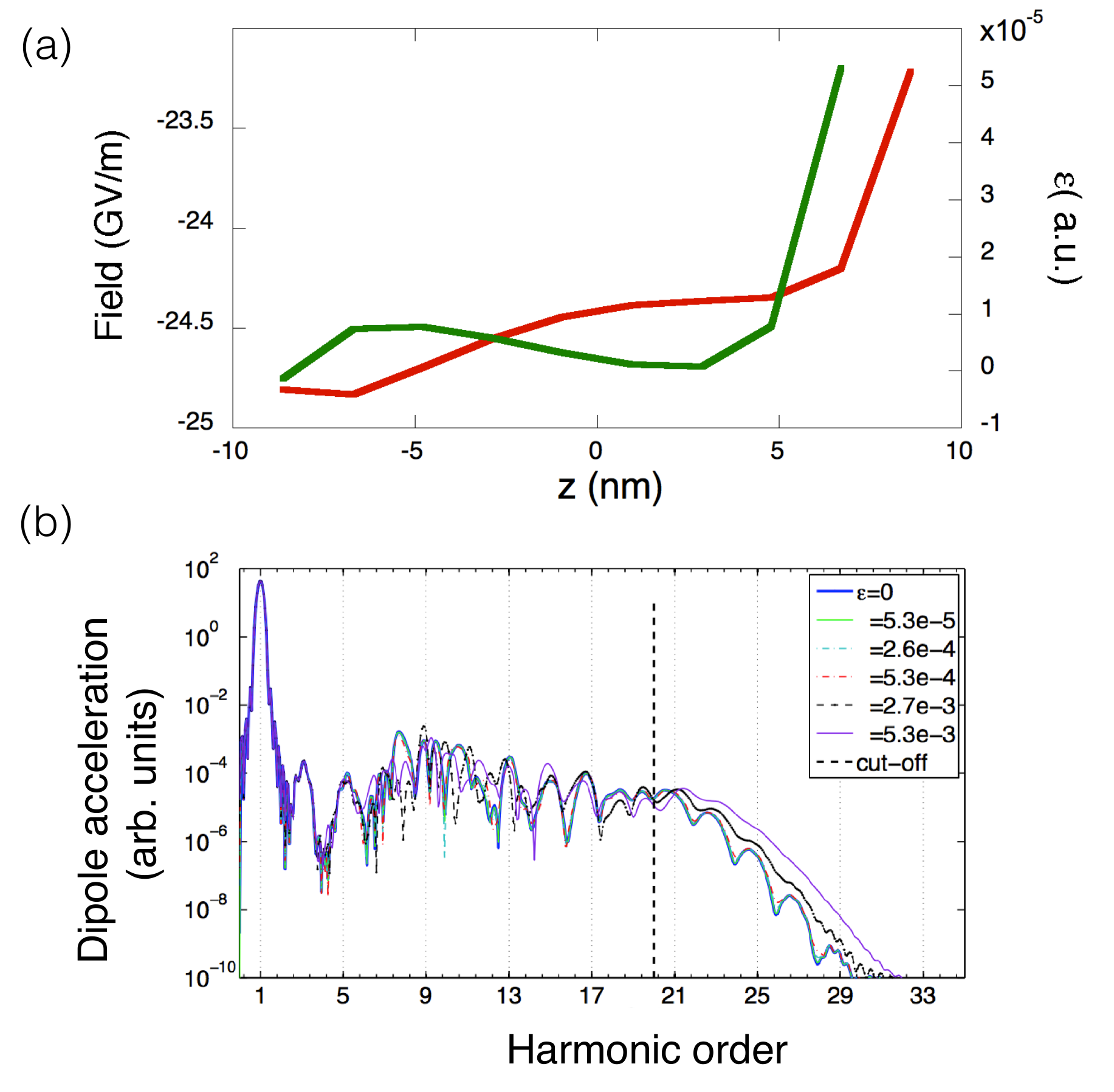}
%\vskip -1 cm 
\caption{(a) Red line: field amplitude profile along the nanoantenna's gap as shown in Fig. 1, at the antenna's center, during the peak of the driving field $(t=t_{max})$ Green line: Inhomogeneity parameter $\epsilon$ (field amplitude gradient in atomic units) along the antenna's gap. (b) Single-atom harmonic spectra generated, blue solid line: without inhomogeneity; green solid line:  with $\epsilon=5.3 \times 10^{-5}$ corresponding to the maximum value shown in (a); and various inhomogeneity parameters above this value. The position of the spectral cut-off for the homogeneous case is shown as a black dashed vertical line.}
\label{fig:inhomogeneity} 
\end{figure}

Our findings demonstrate that phase matching becomes a limit for thicker antennas, due to the intensity variation along the propagation dimension. On the other hand, the softer phase variations along the z axis allow to enhance the harmonic emission by increasing the antenna's gap and, therefore the interaction volume. The increase of the radiating volume can help to raise the contrast between the harmonic emission and the incoherent background \cite{sivis12}.
				
    \section{Conclusions}
        
				In conclusion, we demonstrate that phase matching plays a relevant role in HHG in nanostructures. Despite the nanoscale propagation distances preclude any significant influence of the common sources of phase mismatch, high harmonics have still a substantial phase shift due to the non-perturbative intrinsic phase. Phase-matching at the nanoscale can be used to modify the spatio-temporal properties of the harmonics, through selection of the quantum path trajectories. We also show that phase matching can be engineered modifying the antenna's geometry.
Recent studies of HHG in solids \cite{Osika16} demonstrate the appearance of an intensity-dependent phase, analogous to the intrinsic phase in HHG from atoms. Therefore, our conclusions can also be applied to HHG from crystals in nanostructures \cite{Han16,vampa17}.

    \section*{Appendix: Effect of field's inhomogeneity in the single atom emission}

Figure \ref{fig:inhomogeneity} displays the field amplitude and its gradient along the bow-tie gap distance ($z$ axis) at the instant when the enhanced field reaches its peak ($t_{max}$), for the particular geometry considered in Fig. 1. The field gradient is represented in atomic units, according to the definition of the inhomogeneity parameter $\epsilon$ used in \cite{ciappina12,chacon15}.  The interaction potential in the length gauge reads as $V_{int}=E_h(t) z (1+ \epsilon z/2)$, where $E_h$ is the homogeneous field's amplitude.  	

To find if this field inhomogeneity is capable to introduce modifications in the single-atom harmonic spectrum, we have computed the harmonic emission numerically. This is done integrating the time-dependent Schr\"odinger equation (TDSE), for a model Ar atom interacting with a 4.7 fs FWHM, 800 nm laser pulse with the peak electric field of 24 GV/m and sinus squared envelope. 

We have implemented the numerical solution of the 3D TDSE in cylindrical coordinates, using a Cranck-Nicholson scheme.  We model the Ar atom using the hydrogen-like potential $V(\rho,z)= -Z_{\rm eff} /\sqrt{\rho^2+z^2 }$, in atomic units. The effective charge parameter $Z_{\rm eff}$ is chosen to reproduce the ground state of the Ar atom's ionization potential (0.579 a.u.). We computed this ground state using imaginary time evolution. 
 
We have chosen an inhomogeneity parameter of $\epsilon=5.3 \times 10^{-5}$ a.u., corresponding to the maximum value in Fig. \ref{fig:inhomogeneity}a . The results are depicted in Fig. \ref{fig:inhomogeneity}b, in comparison with the same calculation for an homogeneous field ($\epsilon=0$), and for higher values of the $\epsilon$. The comparison between the different spectra demonstrates  that the the field gradient has no apparent role in the single-atom emission for the interaction parameters considered in our study. We find that the inhomogeneity starts to have its effect on the harmonic spectrum for $\epsilon>5.3 \times 10^{-4}$ a.u., i.e. one order of magnitude above the maximum value in Fig. \ref{fig:inhomogeneity}a.

		\section*{Funding}
		
		Junta de Castilla y Le\'on (SA046U16) and MINECO (FIS2013-44174-P, FIS2016-75652-P, FIS2015-71933-REDT, SEV-2015-0522, FIS2013-46768-P, FIS2016-79508-P).  M. Blanco is funded by FPU grant program of MECD.
A.Chac\'on and M.Lewenstein also acknowledge support from Adv. ERC grant OSYRIS, Generalitat de Catalunya (SGR 874, CERCA Program), and Fundaci\'o Privada Cellex.
    
    \section*{Acknowledgments}

    The authors would like to acknowledge the OSIRIS Consortium, consisting of UCLA and IST (Lisbon, 
    Portugal) for the use of OSIRIS, for providing access to the OSIRIS framework. We thank Antonio Pic\'on for his comments.

%\section*{References}

%Note that letter submissions to \emph{Optica} use an abbreviated reference style. Citations to journal articles should omit the article title and final page number; this abbreviated reference style is produced automatically when the \texttt{$\setminus$setboolean\{shortarticle\}\{true\}} option is selected in the template, if you are using a .bib file for your references. 
%
%However, full references (to aid the editor and reviewers) must be included as well on an informational page that will not count against page length; again this will be produced automatically if you are using a .bib file and have the \texttt{$\setminus$setboolean\{shortarticle\}\{true\}} option selected.

% Bibliography
\bibliography{lplaja_papers}{}

\begin{thebibliography}{10}

\bibitem{popmintchev12}
T.~Popmintchev, M.-C. Chen, D.~Popmintchev, P.~Arpin, S.~Brown, S.~Ali{\v
  s}auskas, G.~Andriukaitis, T.~Bal{\v c}iunas, O.~M{\"u}cke, A.~Pugzlys,
  A.~Baltu{\v s}ka, B.~Shim, S.~E. Schrauth, A.~Gaeta,
  C.~Hern{\'a}ndez-Garc{\'\i}a, L.~Plaja, A.~Becker, A.~Jaron-Becker,
  M.~Murnane, and H.~Kapteyn, ``Bright coherent ultrahigh harmonics in the kev
  x-ray regime from mid-infrared femtosecond lasers,'' {\em Science}, vol.~336,
  no.~6086, pp.~1287--1291, 2012.

\bibitem{paul01}
P.~M. Paul, E.~S. Toma, P.~Breger, G.~Mullot, F.~Aug{\'e}, P.~Balcou, H.~G.
  Muller, and P.~Agostini, ``Observation of a train of attosecond pulses from
  high harmonic generation,'' {\em Science}, vol.~292, no.~5522,
  pp.~1689--1692, 2001.

\bibitem{chini14}
M.~Chini, K.~Zhao, and Z.~Chang, ``{The generation, characterization and
  applications of broadband isolated attosecond pulses},'' {\em Nat. Photon.},
  vol.~8, pp.~178--186, Feb. 2014.

\bibitem{plaja13}
L.~Plaja, R.~Torres, and A.~Za\"ir, eds., {\em Attosecond Physics. Attosecond
  Measurements and Control of Physical Systems}, vol.~177 of {\em Springer
  Series in Optical Sciences}.
\newblock Springer, 2013.

\bibitem{Krausz2009}
F.~Krausz and M.~Ivanov, ``Attosecond physics,'' {\em Rev. Mod. Phys.},
  vol.~81, pp.~163--234, Feb 2009.

\bibitem{kim08}
S.~Kim, J.~Jin, Y.-J. Kim, I.-Y. Park, Y.~Kim, and S.-W. Kim, ``High-harmonic
  generation by resonant plasmon field enhancement,'' {\em Nature}, vol.~453,
  pp.~757--760, June 2008.

\bibitem{sivis12}
M.~Sivis, M.~Duwe, B.~Abel, and C.~Ropers, ``Nanostructure-enhanced atomic line
  emission,'' {\em Nature}, vol.~485, pp.~E1--E3, May 2012.

\bibitem{pfullmann13}
N.~Pfullmann, C.~Waltermann, M.~Noack, S.~Rausch, T.~Nagy, C.~Reinhardt,
  M.~Kova\v{c}ev, V.~Knittel, R.~Bratschitsch, D.~Akemeier, A.~H\"utten,
  A.~Leitenstorfer, and U.~Morgner, ``Bow-tie nano-antenna assisted generation
  of extreme ultraviolet radiation,'' {\em New J. Phys.}, vol.~15, no.~9,
  p.~093027, 2013.

\bibitem{Han16}
S.~Han, H.~Kim, Y.~W. Kim, Y.-J. Kim, S.~Kim, I.-Y. Park, and S.-W. Kim,
  ``High-harmonic generation by field enhanced femtosecond pulses in
  metal-sapphire nanostructure,'' {\em Nat. Commun.}, vol.~7, p.~13105, 2016.

\bibitem{vampa17}
G.~Vampa, B.~G. Ghamsari, S.~Siadat~Mousavi, T.~J. Hammond, A.~Olivieri,
  E.~Lisicka-Skrek, A.~Y. Naumov, D.~M. Villeneuve, A.~Staudte, P.~Berini, and
  P.~B. Corkum, ``{Plasmon-enhanced high-harmonic generation from silicon},''
  {\em Nat. Phys.}, vol.~64, pp.~39--5, Apr. 2017.

\bibitem{husakou11}
A.~Husakou, S.-J. Im, and J.~Herrmann, ``Theory of plasmon-enhanced high-order
  harmonic generation in the vicinity of metal nanostructures in noble gases,''
  {\em Phys. Rev. A}, vol.~83, p.~043839, Apr 2011.

\bibitem{stebbings11}
S.~L. Stebbings, F.~S. mann, Y.-Y. Yang, A.~Scrinzi, M.~Durach, A.~Rusina,
  M.~I. Stockman, and M.~F. Kling, ``Generation of isolated attosecond extreme
  ultraviolet pulses employing nanoplasmonic field enhancement: optimization of
  coupled ellipsoids,'' {\em New J. Phys.}, vol.~13, no.~7, p.~073010, 2011.

\bibitem{choi12}
J.~Choi, S.~Kim, I.-Y. Park, D.-H. Lee, S.~Han, and S.-W. Kim, ``Generation of
  isolated attosecond pulses using a plasmonic funnel-waveguide,'' {\em New J.
  Phys.}, vol.~14, no.~10, p.~103038, 2012.

\bibitem{ciappina16}
M.~F. Ciappina, J.~A. P\'erez-Hern\'andez, A.~S. Landsman, W.~A. Okell,
  S.~Zherebtsov, B.~F\"org, J.~Sch\"oz, L.~Seiffert, T.~Fennel, T.~Shaaran,
  T.~Zimmermann, A.~Chac\'on, R.~Guichard, A.~Za\"ir, J.~W.~G. Tisch, J.~P.
  Marangos, T.~Witting, A.~Braun, S.~A. Maier, L.~Roso, M.~Kr\"uger,
  P.~Hommelhoff, M.~F. Kling, F.~Krausz, and M.~Lewenstein, ``Attosecond
  physics at the nanoscale,'' {\em Rep. Prog. Phys.}, vol.~80, no.~5,
  p.~054401, 2017.

\bibitem{summers14}
A.~M. Summers, A.~S. Ramm, G.~Paneru, M.~F. Kling, B.~N. Flanders, and C.~A.
  Trallero-Herrero, ``Optical damage threshold of au nanowires in strong
  femtosecond laser fields,'' {\em Opt. Express}, vol.~22, pp.~4235--4246, Feb
  2014.

\bibitem{kim12}
S.~Kim, J.~Jin, Y.-J. Kim, I.-Y. Park, Y.~Kim, and S.-W. Kim, ``{Kim et al.
  reply},'' {\em Nature}, vol.~485, pp.~E1--E3, May 2012.

\bibitem{sivis13}
M.~Sivis, M.~Duwe, B.~Abel, and C.~Ropers, ``{Extreme-ultraviolet light
  generation in plasmonic nanostructures},'' {\em Nat. Phys.}, vol.~9,
  pp.~304--309, Mar. 2013.

\bibitem{gaarde08}
M.~B. Gaarde, J.~L. Tate, and K.~J. Schafer, ``Macroscopic aspects of
  attosecond pulse generation,'' {\em J. Phys. B}, vol.~41, no.~13, p.~132001,
  2008.

\bibitem{hernandez16}
C.~Hern\'andez-Garc\'ia, T.~Popmintchev, M.~M. Murnane, H.~C. Kapteyn,
  L.~Plaja, A.~Becker, and A.~Jaron-Becker, ``Group velocity matching in
  high-order harmonic generation driven by mid-infrared lasers,'' {\em New J.
  Phys.}, vol.~18, no.~7, p.~073031, 2016.

\bibitem{hernandez15}
C.~Hern\'{a}ndez-Garc\'{i}a, W.~Holgado, L.~Plaja, B.~Alonso, F.~Silva,
  M.~Miranda, H.~Crespo, and I.~J. Sola, ``Carrier-envelope-phase insensitivity
  in high-order harmonic generation driven by few-cycle laser pulses,'' {\em
  Opt. Express}, vol.~23, pp.~21497--21508, Aug 2015.

\bibitem{schafer93}
K.~J. Schafer, B.~Yang, L.~F. DiMauro, and K.~C. Kulander, ``Above threshold
  ionization beyond the high harmonic cutoff,'' {\em Phys. Rev. Lett.},
  vol.~70, pp.~1599--1602, Mar 1993.

\bibitem{corkum93}
P.~B. Corkum, ``Plasma perspective on strong field multiphoton ionization,''
  {\em Phys. Rev. Lett.}, vol.~71, pp.~1994--1997, Sep 1993.

\bibitem{lewenstein94}
M.~Lewenstein, P.~Balcou, M.~Y. Ivanov, A.~L'Huillier, and P.~B. Corkum,
  ``Theory of high-harmonic generation by low-frequency laser fields,'' {\em
  Phys. Rev. A}, vol.~49, pp.~2117--2132, Mar 1994.

\bibitem{lewestein95}
M.~Lewenstein, P.~Sali\`eres, and A.~L'Huillier, ``Phase of the atomic
  polarization in high-order harmonic generation,'' {\em Phys. Rev. A},
  vol.~52, pp.~4747--4754, Dec 1995.

\bibitem{hernandez13}
C.~Hern\'andez-Garc\'{\i}a, I.~J. Sola, and L.~Plaja, ``Signature of the
  transversal coherence length in high-order harmonic generation,'' {\em Phys.
  Rev. A}, vol.~88, p.~043848, Oct 2013.

\bibitem{salieres01}
P.~Sali{\`e}res, B.~Carr{\'e}, L.~Le~D{\'e}roff, F.~Grasbon, G.~G. Paulus,
  H.~Walther, R.~Kopold, W.~Becker, D.~B. Milo{\v s}evi{\'c}, A.~Sanpera, and
  M.~Lewenstein, ``Feynman{\textquoteright}s path-integral approach for
  intense-laser-atom interactions,'' {\em Science}, vol.~292, no.~5518,
  pp.~902--905, 2001.

\bibitem{mairesse03}
Y.~Mairesse, A.~de~Bohan, L.~J. Frasinski, H.~Merdji, L.~C. Dinu,
  P.~Monchicourt, P.~Breger, M.~Kova{\v c}ev, R.~Ta{\"\i}eb, B.~Carr{\'e},
  H.~G. Muller, P.~Agostini, and P.~Sali{\`e}res, ``Attosecond synchronization
  of high-harmonic soft x-rays,'' {\em Science}, vol.~302, no.~5650,
  pp.~1540--1543, 2003.

\bibitem{zair08}
A.~Za\"{\i}r, M.~Holler, A.~Guandalini, F.~Schapper, J.~Biegert, L.~Gallmann,
  U.~Keller, A.~S. Wyatt, A.~Monmayrant, I.~A. Walmsley, E.~Cormier,
  T.~Auguste, J.~P. Caumes, and P.~Sali\`eres, ``Quantum path interferences in
  high-order harmonic generation,'' {\em Phys. Rev. Lett.}, vol.~100,
  p.~143902, Apr 2008.

\bibitem{hernandez12}
C.~Hern\'andez-Garc\'ia and L.~Plaja, ``Off-axis compensation of attosecond
  pulse chirp,'' {\em J. Phys. B-At. Mol. Opt.}, vol.~45, no.~7, p.~074021,
  2012.

\bibitem{hernandez13Z}
C.~Hern\'andez-Garc\'{\i}a, J.~A. P\'erez-Hern\'andez, T.~Popmintchev, M.~M.
  Murnane, H.~C. Kapteyn, A.~Jaron-Becker, A.~Becker, and L.~Plaja,
  ``Zeptosecond high harmonic kev x-ray waveforms driven by midinfrared laser
  pulses,'' {\em Phys. Rev. Lett.}, vol.~111, p.~033002, Jul 2013.

\bibitem{hernandez15Q}
C.~Hern\'andez-Garc\'ia, J.~S. Rom\'an, L.~Plaja, and A.~Pic\'on,
  ``Quantum-path signatures in attosecond helical beams driven by optical
  vortices,'' {\em New J. Phys.}, vol.~17, no.~9, p.~093029, 2015.

\bibitem{fonseca02}
R.~A. Fonseca, L.~O. Silva, F.~S. Tsung, V.~K. Decyk, W.~Lu, C.~Ren, W.~B.
  Mori, S.~Deng, S.~Lee, T.~Katsouleas, and J.~C. Adam, {\em OSIRIS: A
  Three-Dimensional, Fully Relativistic Particle in Cell Code for Modeling
  Plasma Based Accelerators}, pp.~342--351.
\newblock Springer, 2002.

\bibitem{fonseca2008}
R.~A. Fonseca, S.~F. Martins, L.~O. Silva, J.~W. Tonge, F.~S. Tsung, and W.~B.
  Mori, ``One-to-one direct modeling of experiments and astrophysical
  scenarios: pushing the envelope on kinetic plasma simulations,'' {\em Plasma
  Phys. Control. Fusion}, vol.~50, no.~12, p.~124034, 2008.

\bibitem{fonseca2013}
R.~A. Fonseca, J.~Vieira, F.~Fiuza, A.~Davidson, F.~S. Tsung, W.~B. Mori, and
  L.~O. Silva, ``Exploiting multi-scale parallelism for large scale numerical
  modelling of laser wakefield accelerators,'' {\em Plasma Phys. Control.
  Fusion}, vol.~55, no.~12, p.~124011, 2013.

\bibitem{ciappina12}
M.~F. Ciappina, J.~Biegert, R.~Quidant, and M.~Lewenstein,
  ``High-order-harmonic generation from inhomogeneous fields,'' {\em Phys. Rev.
  A}, vol.~85, p.~033828, Mar 2012.

\bibitem{hernandez10}
C.~Hern\'andez-Garc\'{\i}a, J.~A. P\'erez-Hern\'andez, J.~Ramos, E.~C. Jarque,
  L.~Roso, and L.~Plaja, ``High-order harmonic propagation in gases within the
  discrete dipole approximation,'' {\em Phys. Rev. A}, vol.~82, p.~033432, Sep
  2010.

\bibitem{keldysh64}
L.~V. Keldysh, ``Ionization in the field of a strong electromagnetic wave,''
  {\em Zh. Eksp. Teor. Fiz.}, vol.~47, p.~1945, 1965.

\bibitem{faisal73}
F.~H.~M. Faisal, ``Multiple absorption of laser photons by atoms,'' {\em J.
  Phys. B-At. Mol. Opt.}, vol.~6, no.~4, p.~L89, 1973.

\bibitem{reiss80}
H.~R. Reiss, ``Effect of an intense electromagnetic field on a weakly bound
  system,'' {\em Phys. Rev. A}, vol.~22, pp.~1786--1813, Nov 1980.

\bibitem{becker97}
W.~Becker, A.~Lohr, M.~Kleber, and M.~Lewenstein, ``A unified theory of
  high-harmonic generation: Application to polarization properties of the
  harmonics,'' {\em Phys. Rev. A}, vol.~56, pp.~645--656, Jul 1997.

\bibitem{perez09}
J.~A. P\'{e}rez-Hern\'{a}ndez, L.~Roso, and L.~Plaja, ``Harmonic generation
  beyond the strong-field approximation: the physics behind the
  short-wave-infrared scaling laws,'' {\em Opt. Express}, vol.~17,
  pp.~9891--9903, Jun 2009.

\bibitem{perez11}
J.~A. P\'erez-Hern\'andez, C.~Hern\'andez-Garc\'ia, J.~Ramos, E.~C. Jarque,
  L.~Plaja, and L.~Roso, {\em New Methods For Computing High-Order Harmonic
  Generation and Propagation}, pp.~145--162.
\newblock Springer, 2011.

\bibitem{Birdsall:1991}
C.~K. Birdsall and A.~B. Langdon, {\em Plasma Physics via Computer Simulation}.
\newblock IOP Publishing Ltd., 1991.

\bibitem{antoine95}
P.~Antoine, B.~Piraux, and A.~Maquet, ``Time profile of harmonics generated by
  a single atom in a strong electromagnetic field,'' {\em Phys. Rev. A},
  vol.~51, pp.~R1750--R1753, Mar 1995.

\bibitem{Sansone2006}
G.~Sansone, E.~Benedetti, F.~Calegari, C.~Vozzi, L.~Avaldi, R.~Flammini,
  L.~Poletto, P.~Villoresi, C.~Altucci, R.~Velotta, S.~Stagira,
  S.~De~Silvestri, and M.~Nisoli, ``Isolated single-cycle attosecond pulses,''
  {\em Science}, vol.~314, no.~5798, pp.~443--446, 2006.

\bibitem{Osika16}
E.~N. Osika, A.~Chac\'on, L.~Ortmann, N.~Su\'arez, J.~A. P\'erez-Hern\'andez,
  B.~Szafran, M.~F. Ciappina, F.~Sols, A.~S. Landsman, and M.~Lewenstein,
  ``Wannier-bloch approach to localization in high-harmonics generation in
  solids,'' {\em Phys. Rev. X}, vol.~7, p.~021017, May 2017.

\bibitem{chacon15}
A.~Chac\'on, M.~F. Ciappina, and M.~Lewenstein, ``Numerical studies of
  light-matter interaction driven by plasmonic fields: The velocity gauge,''
  {\em Phys. Rev. A}, vol.~92, p.~063834, Dec 2015.

\end{thebibliography}
\bibliographystyle{ieeetr} 

% Full bibliography added automatically for Optics Letters submissions
% Note that this extra page will not count against page length

%Manual citation list
%\begin{thebibliography}{1}
%\bibitem{Zhang:14}
%Y.~Zhang, S.~Qiao, L.~Sun, Q.~W. Shi, W.~Huang, %L.~Li, and Z.~Yang,
 % \enquote{Photoinduced active terahertz metamaterials with nanostructured
  %vanadium dioxide film deposited by sol-gel method,} Opt. Express \textbf{22},
  %11070--11078 (2014).
% \end{thebibliography}

\end{document}